\documentclass{nature}

\usepackage{graphicx}
\usepackage{siunitx}
\usepackage{amsmath, mathtools, bm}
\usepackage{amssymb,amsfonts} 
\usepackage{stmaryrd}
\usepackage{tipa}

\usepackage[colorlinks,citecolor=blue,linkcolor = blue,urlcolor=blue,bookmarks=false,hypertexnames=true]{hyperref}

\makeatletter
\let\saved@includegraphics\includegraphics
\AtBeginDocument{\let\includegraphics\saved@includegraphics}
\renewenvironment*{figure}{\@float{figure}}{\end@float}
\makeatother

\newcommand{\ket}[1]{\ensuremath{|#1\rangle}}

\title{Pauli blocking of stimulated emission in a degenerate Fermi gas}

\author{R. Jannin$^{1}$, Y. van der Werf$^{1}$, K. Steinebach$^{1}$, H. L. Bethlem$^{1}$ \& K. S. E. Eikema$^{1}$}

\begin{document}

\maketitle

\begin{affiliations}
 \item LaserLab, Department of Physics and Astronomy, Vrije Universiteit, De Boelelaan 1081, 1081 HV Amsterdam, The Netherlands
\end{affiliations}

\begin{abstract}
The Pauli exclusion principle in quantum mechanics has a profound influence on the structure of matter and on interactions between fermions. Almost 30 years ago it was predicted that the Pauli exclusion principle could lead to a suppression of spontaneous emission, and only recently several experiments \cite{Deb2021, Sanner2021, Margalit2021} confirmed this phenomenon. 
Here we report that this so-called Pauli blockade not only affects incoherent processes but also, more generally, coherently driven systems.
It manifests itself as an intriguing sub-Doppler narrowing of a doubly-forbidden transition profile in an optically trapped Fermi gas of $^3\mathrm{He}$. By actively pumping atoms out of the excited state, we break the coherence of the excitation and lift the narrowing effect, confirming the influence of Pauli blockade on the transition profile. This new insight into the interplay between quantum statistics and coherent driving is a promising development for future applications involving fermionic systems.
\end{abstract}

The Pauli exclusion principle arises from the simple requirement of antisymmetrization of two-fermion wavefunctions under particle exchange \cite{Pauli1925}, but it has a remarkable impact on how our world is shaped. It determines the electronic structure of atoms, which dictates their chemical properties, and it gives rise to the Fermi energy of electron ensembles in solid state materials, leading to electrical (semi-)conductivity. Another example is the Fermi pressure which stabilizes the densest observable matter in our universe like white dwarves and neutron stars against gravitational collapse. In the field of ultracold atomic physics, the Pauli exclusion principle has a direct impact on the collisional properties of identical fermions, due to the absence of $s$-wave (even parity) collisions, which has been observed experimentally \cite{Gupta2003, Roati2004}, and has cleared the way towards unprecedented fractional uncertainties at or below the $10^{-18}$ level of state-of-the-art fermionic 1D lattice clocks \cite{Yamanaka2015, Tyumenev2016, Campbell2017, Bothwell2022, Zheng2022}. 

Since the idea was suggested almost three decades ago that spontaneous decay of an excited ultracold fermion confined in a Fermi sea would be suppressed due to quantum statistics, it has regularly attracted theoretical interest \cite{Helmerson1990, Imamoglu1994, Javanainen1995, Busch1998, DeMarco1998, Shuve2009, OSullivan2009, Sandner2011}. 
This suppression of spontaneous emission can become relevant in photon scattering events, where the absorption and subsequent spontaneous emission of a photon imparts a momentum transfer $\hbar k = \hbar(k_\mathrm{abs}-k_\mathrm{emi})$ on the atom. 
If the imparted photon recoil is smaller than the Fermi momentum of the Fermi sea, these scattering events couple to states that are already occupied, and are thus strongly suppressed. This so-called Pauli blockade of spontaneous emission in ultracold degenerate Fermi gases has only recently been experimentally observed \cite{Sanner2021, Deb2021, Margalit2021}.

In this work we demonstrate, for the first time to our knowledge, the Pauli blockade of \textit{stimulated} emission, which leads to the intriguing phenomenon of narrowing the linewidth of the $2^3\mathrm{S}_1 \rightarrow 2^1\mathrm{S}_0$ transition that is studied in our experiment.
This transition at \SI{1557}{nm} connects the two metastable states of helium. Due to the very small Einstein coefficient of about $9\times 10^{-8}\ \mathrm{s}^{-1}$, the upper state lifetime fully determines the natural linewidth of \SI{8}{Hz}, making this an ideal candidate for precision spectroscopy in helium. Moreover, it is trivial to make the coherent driving Rabi frequency ($2\pi\times 40\ \mathrm{Hz}$ for our experimental conditions) orders of magnitude larger than the Einstein coefficient, eliminating all influence of spontaneous decay back to the $2^3\mathrm{S}_1$ state.

\begin{figure}[ht]
\center
\includegraphics[scale=0.45]{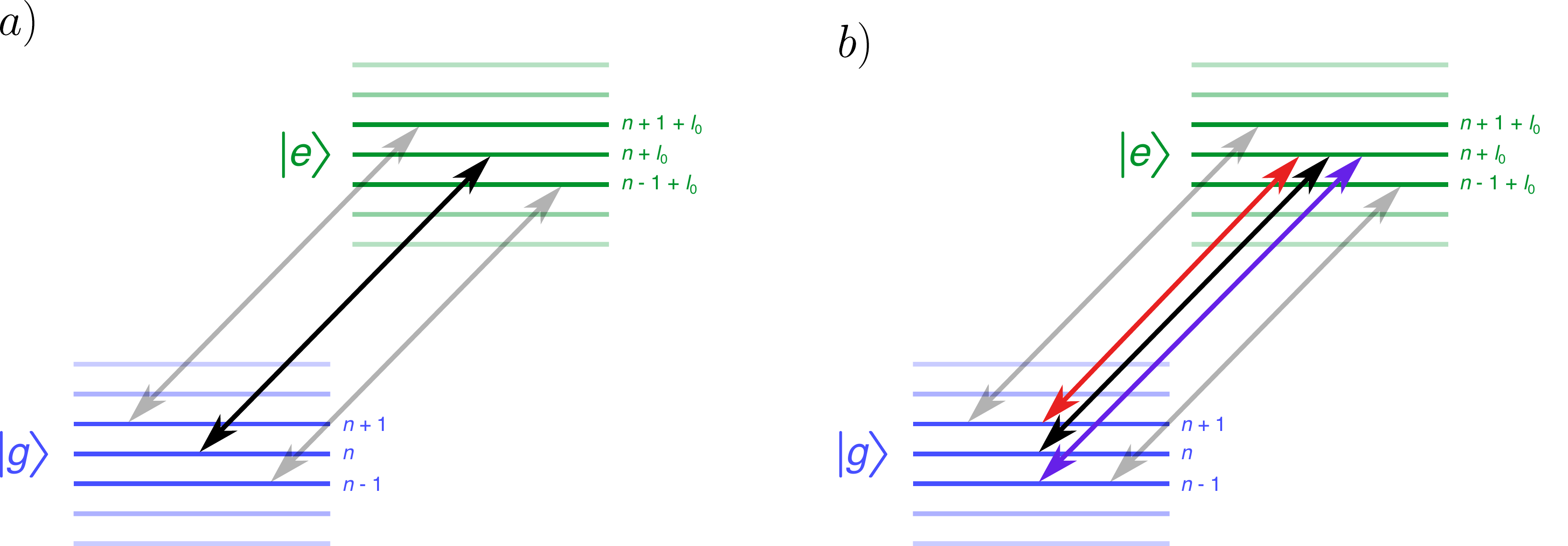}
\caption{Couplings between the magic wavelength optical dipole trap motional states, induced by the excitation light. (a) When the excitation light contains a single frequency, only pairs of motional states with a fixed energy difference (equal to $\ell_0$ vibrational quanta) are coupled (carrier transitions). (b) When the spectrum of the excitation laser is broader than the energy splitting of the vibrational states, several other transitions are possible too (sideband transitions).
}
\label{fig:carriers_sidebands}
\end{figure}

The excitation is performed in a degenerate Fermi gas of $^3\mathrm{He}$ confined in a crossed optical dipole trap (ODT) at the magic wavelength (where the trapping potential is identical for both $2^3\mathrm{S}_1$ and $2^1\mathrm{S}_0$ states). 
If the linewidth of the excitation source is narrow enough to resolve the energy difference between the motional states induced by the trapping potential, only pairs of states with the same energy difference are coupled, as illustrated in Fig.\ref{fig:carriers_sidebands}(a). We define this as carrier transitions, and the hole left in the lower state by excitation to the upper state can only be refilled by stimulated emission of the same atom back to the lower state again. The evolution of the system can be described as an ensemble of independent atoms each performing its own Rabi oscillation. The absorption profile will therefore simply reflect the momentum distribution of the atoms in the trap, in the form of a Doppler broadening.

If, on the other hand, the linewidth of the excitation laser is much broader than the energy spacing of the motional levels (which is the case in our experiment), each vibrational state couples to multiple others. This situation is depicted in Fig.\ref{fig:carriers_sidebands}(b).
The sideband transitions lead to an exchange of motional states, which is affected by the Pauli exclusion principle.
Upon absorption of a photon, de-excitation is again only possible towards states which are not occupied, but now the laser bandwidth covers many more states. The excitation profile will then reflect both the absorption profile, determined by the phase space density, and stimulated emission downward again, which depends on the distribution of holes in the Fermi gas. It ultimately leads to a narrowing of the measured transition, which counter-intuitively happens only if the laser is broad enough to couple different motional states. The model explaining this narrowing effect will be developed in the next paragraphs.

We can describe the excitation process by the following Hamiltonian in the interaction picture:
\begin{equation}
H = \sum_{nm} a_{m-n}\dfrac{\Omega_{g_n,e_m}}{2}\left( \hat{e}_{m}^\dagger\hat{g}_{n} + \hat{g}_{n}^\dagger\hat{e}_{m} \right) \equiv \sum_{nm} H_{g_n e_m} \left( \hat{e}_{m}^\dagger\hat{g}_{n} + \hat{g}_{n}^\dagger\hat{e}_{m} \right) ,
\label{eq:hamiltonian}
\end{equation}
where $\hat{g}_n^\dagger$ and $\hat{e}_m^\dagger$ ($\hat{g}_n$ and $\hat{e}_m$) represent the creation (annihilation) operators of a fermion in state $\ket{g,n}$ and $\ket{e,m}$ respectively, where $n$ and $m$ represent the motional quantum states for atoms in the lower ($\ket{g}$) and upper ($\ket{e}$) internal atomic states (see Fig.\ref{fig:carriers_sidebands}), and $\Omega_{g_n,e_m}$ represents the Rabi frequencies coupling these states \cite{Leibfried2003} (see the Supplementary material for more details). 
The weights $a_{m-n}$ in Eq.\eqref{eq:hamiltonian} describe the spectral intensity distribution of the excitation laser light, which is peaked at $|m-n| = \ell_0$. When e.g.~only $a_{\ell_0}$ is non-zero, then only carrier transitions are driven, as shown in Fig.\ref{fig:carriers_sidebands}(a). 
In order to capture the physics resulting from the excitation light, we compute the transition rates $\Gamma_{g_n\rightarrow e_m}$ and $\Gamma_{e_m\rightarrow g_n}$ using Fermi's golden rule (see the Supplementary material for details):
\begin{align}
\Gamma_{g_n\rightarrow e_m} & = \dfrac{2\pi}{\hbar^2\omega_g} \left| H_{g_n e_m} \right|^2n_{g_n} \left( 1-n_{e_m} \right) \mathrm{\ and} \label{eq:rate_ge}\\
\Gamma_{e_m \rightarrow g_n} & = \dfrac{2\pi}{\hbar^2\omega_e} \left| H_{g_n e_m} \right|^2 n_{e_m} \left( 1-n_{g_n} \right),\label{eq:rate_eg}
\end{align}
with $n_{g_n}$ and $n_{e_m}$ representing the occupation numbers of the different states and $\omega_g$ ($\omega_e$) corresponding to the trapping frequency felt by the internal state $\ket{g}$ ($\ket{e}$) (for a magic wavelength ODT, $\omega_g = \omega_e$). These excitation rates reflect the Pauli exclusion principle since they both depend on the occupation of the initial state and the availability of holes in the final state.
From this we obtain the shape of our spectroscopy signal (measured as depletion of the $2^3\mathrm{S}_1$ population):
\begin{equation}
\mathcal{S} \propto \sum_{nm} \left|a_{m-n}\right|^2\left| \Omega_{g_n e_m} \right|^2 \left(n_{g_n}(0) -\dfrac{\sum_k \left|a_{m-k}\right|^2\left| \tilde{\Omega}_{g_k e_m}\right|^2  n_{g_k}(0)\left( 1-n_{g_n}(0) \right)}{\sum_k \left|a_{m-k}\right|^2\left| \tilde{\Omega}_{g_k e_m}\right|^2 + f_g \Gamma_0/\Omega^2} \right),
\label{eq:real_model}
\end{equation}
in which $1/\Gamma_0$ represents the lifetime of the $\ket{e}$ state, $f_g=\omega_g/2\pi$ and $\tilde{\Omega}_{g_k e_m}=\Omega_{g_k e_m}/\Omega$ (see the Supplementary material for an elaborate derivation). The first term describes the contribution of absorption to the signal and the second one accounts for exchange of motional states within the $\ket{g}$ manifold through stimulated emission, which is subject to Pauli blocking.

To get to a practically applicable model, the motional states are treated in a semi-classical approach and denoted with $|\mathbf{r}, \mathbf{k}\rangle$, where $\mathbf{r}$ represents the position and $\hbar\mathbf{k}$ the momentum of the atoms. The number density of the initial Fermi gas is given by the Fermi-Dirac distribution function \cite{Butts1997, Schneider1998}:
\begin{equation}
\rho_\mathrm{g}(\mathbf{r}, \mathbf{k}) = \dfrac{1}{(2\pi)^3}\dfrac{1}{1+\exp\left( \beta H_\mathrm{g}(\mathbf{r}, \mathbf{k}) - \beta\mu\right)},
\end{equation}
with $\beta = 1/k_\mathrm{B}T$, where $T$ and $\mu$ are the temperature and chemical potential of the gas respectively, and $H_g(\mathbf{r}, \mathbf{k})$ is the Hamiltonian describing state $|g\rangle$. 

Similarly to References \cite{Busch1998, Shuve2009}, we use a local density approach to estimate the effective excitation rate between $|g\rangle$ and $| e \rangle$. 
Expression \eqref{eq:real_model} yields (see the Supplementary material for details):
\begin{equation}
\mathcal{S}(\omega) \propto \mathcal{S}_\mathrm{Absorption}(\omega)\left( 1 - \mathcal{M}(\omega) \right),
\label{eq:general_S}
\end{equation}
with 
\begin{equation}
\mathcal{S}_\mathrm{Absorption}(\omega) \propto \int \mathrm{d}^3 \mathbf{r} \int \mathrm{d}^3 \mathbf{k}\ \rho_\mathrm{g}(\mathbf{r}, \mathbf{k}) \delta(\omega - \omega_{\mathbf{r}, \mathbf{k}}),
\label{eq:Doppler}
\end{equation}
and a modification factor $\mathcal{M}(\omega)$, representing the Pauli-blocked stimulated emission, defined as:
\begin{equation}
\mathcal{M}(\omega) = \dfrac{\int \mathrm{d}^3 \mathbf{r} \int \mathrm{d}^3 \mathbf{k}\ \rho_\mathrm{g}(\mathbf{r}, \mathbf{k}) \left( 1 - \rho_\mathrm{g}(\mathbf{r}, \mathbf{k})\right) \delta(\omega - \omega_{\mathbf{r}, \mathbf{k}}) }{\int \mathrm{d}^3 \mathbf{r} \int \mathrm{d}^3 \mathbf{k}\ \rho_\mathrm{g}(\mathbf{r}, \mathbf{k}) \delta(\omega - \omega_{\mathbf{r}, \mathbf{k}})}.
\label{eq:M}
\end{equation}

In the Supplementary material a full evaluation of equations \eqref{eq:general_S}-\eqref{eq:M} is given. In principle $\mathcal{M}(\omega)$ depends also on the spectral profile of the excitation laser, but to include this effect would significantly complicate the calculation (see the Supplementary). As the linewidth of the excitation laser is much broader than the energy spacing of the motional states, we  therefore only consider that the main contribution will come from the center frequency and thus neglect the small change in position or momentum.

\begin{figure}[ht]
\center
\includegraphics[scale=0.315]{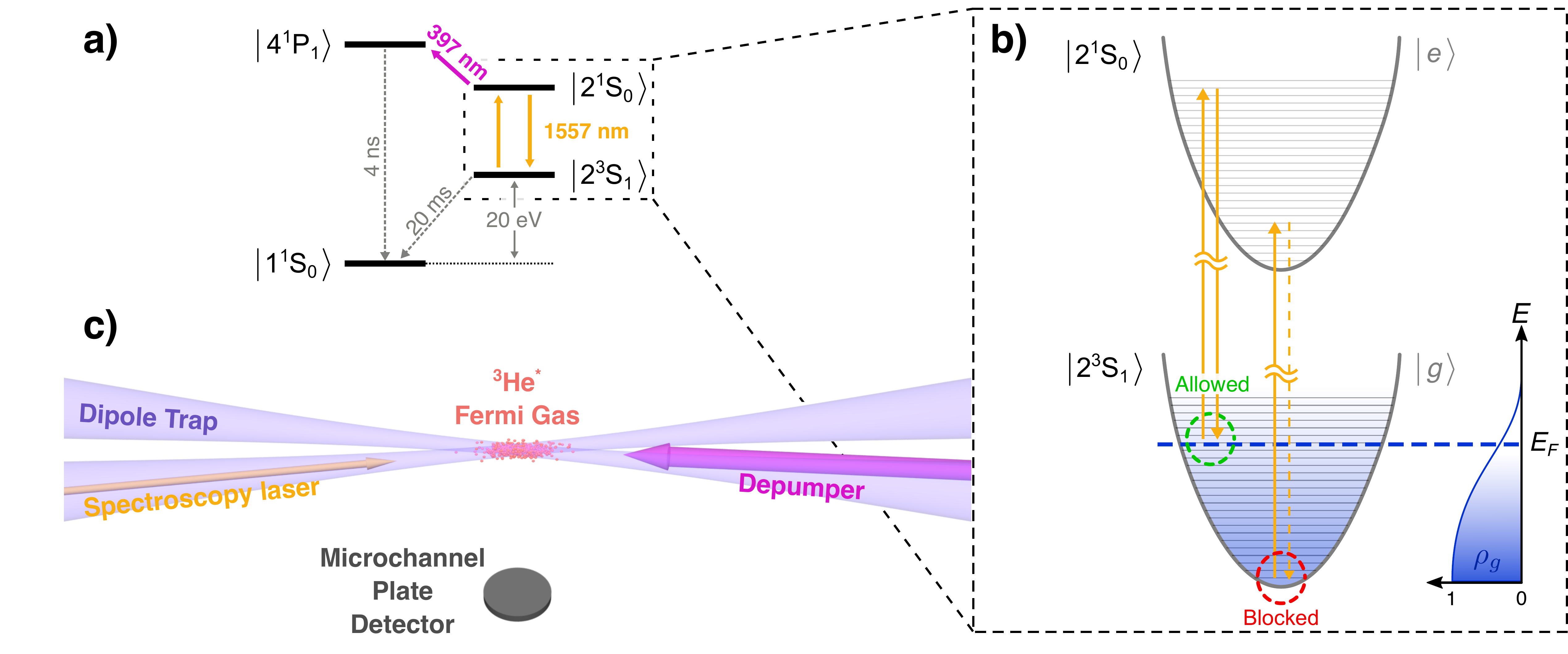}
\caption{(a) Energy scheme showing the relevant electronic levels of helium. A depumper between the $2^1\mathrm{S}_0$ and the $4^1\mathrm{P}_1$ states can artificially reduce the lifetime of the $2^1\mathrm{S}_0$ state. 
(b) Vibrational states induced by the optical dipole potential and depiction of the Pauli blocking mechanism depending on the state occupation: stimulated emission is inhibited for low-energy states whereas it is allowed for high-energy states. This results in a narrowing of the spectrum.  (c) Schematic of the experiment. The Fermi gas is trapped in a crossed dipole trap at \SI{320}{nm}. The spectroscopy beam at \SI{1557}{nm} counterpropagates with respect to one of the beams from the ODT. A depumper at \SI{397}{nm} is sent on the atomic cloud along its long axis. After excitation, the remaining part of the DFG is released from the ODT and is detected by a microchannel plate detector located below the trapping region.}
\label{fig:levels_flopping}
\end{figure}

Figure \ref{fig:levels_flopping}(a) shows the energy levels of helium that are relevant to our experiment.
A graphical representation of the influence of the ODT induced potentials and of $\mathcal{M}(\omega)$ on the $2^3\mathrm{S}_1\rightarrow 2^1\mathrm{S}_0$ excitation process is also shown in Fig.\ref{fig:levels_flopping}(b), illustrating the narrowing mechanism. Atoms with (low) momentum $p\ll p_F$ ($p_F$ representing the Fermi momentum) are excited at the center of the spectral profile, corresponding to atoms populating the low energy states of the confining potential. For those atoms almost no holes are available in a degenerate Fermi gas to go back to, so $\mathcal{M}(\omega)$ goes to zero and stimulated emission is suppressed. On the other hand, atoms with $p\simeq p_F$ are excited at the wings of the spectral profile due to the Doppler effect, for these atoms the availability of holes goes to unity. Stimulated emission back is allowed, leading to a reduced effective excitation rate from $\ket{g}$ to $\ket{e}$. As the Doppler effect is the only broadening mechanism for fermions in a magic wavelength ODT, the spectral linewidth narrows compared to pure Doppler broadening. Moreover, the effect does not induce any additional shift on the center frequency since it is fully symmetrical in momentum space. The validity of equation \eqref{eq:general_S} describing this phenomenon was confirmed by also performing a few-body simulation based on numerically solving the master equation describing the dynamics of the system (see the Supplementary material).

Most of the experimental apparatus has been described in earlier works \cite{Vassen2008, vanRooij2011, Rengelink2018}.
To produce a degenerate Fermi gas (DFG) of $^3\mathrm{He}$ in the metastable state $2^3\mathrm{S}_1(F=3/2, m_F=+3/2)$ we first perform sympathetic cooling along with bosonic $^4\mathrm{He}$. The mixture is evaporatively cooled down to quantum degeneracy in a cloverleaf magnetic trap until only a pure Fermi gas of $^3\mathrm{He}$ is left. The gas is then transferred to a crossed optical dipole trap (ODT) at the magic wavelength for the $2^3\mathrm{S}_1\rightarrow 2^1\mathrm{S}_0$ transition at \SI{319.8}{nm} \cite{Notermans2014, Rengelink2016}, with typical confining frequencies of $\omega_\sslash =2\pi\times$\SI{15}{Hz} in the axial direction and $\omega_\perp =2\pi\times$\SI{120}{Hz} in the radial ones. We typically produce Fermi gases composed of a few $10^5$ to few $10^6$ atoms at temperatures $T/T_\mathrm{F}$ ranging from 1.5 to 0.25. The excitation to the $2^1\mathrm{S}_0(F=1/2, m_F=+1/2)$ is then performed using a laser beam at \SI{1557}{nm}, with a spectral linewidth of about \SI{5}{kHz} \cite{Rengelink2018}, counter-propagating with respect to the incident beam of the ODT, corresponding to the axial direction of the trap, as shown in Fig.\ref{fig:levels_flopping}(c). After 3 seconds of excitation, the metastable helium cloud is released from the trap and falls under gravity onto a micro-channel plate detector (MCP) producing a time-of-flight signal. This provides a way to measure the number of atoms of the gas, its temperature and chemical potential. Because of the relatively short lifetime of about \SI{20}{ms} of the $2^1\mathrm{S}_0$ state compared to the time of free-fall, only atoms left in the $2^3\mathrm{S}_1$ state are detected, and excitation to the singlet metastable state is translated into a reduction of the number of trapped atoms. This is our spectroscopic signal, and each realization of the experimental sequence is followed by one where a measurement of the DFG atom number without excitation light is made to probe the fluctuations of the atom number over time.

\begin{figure}[ht]
\center
\includegraphics[scale=1]{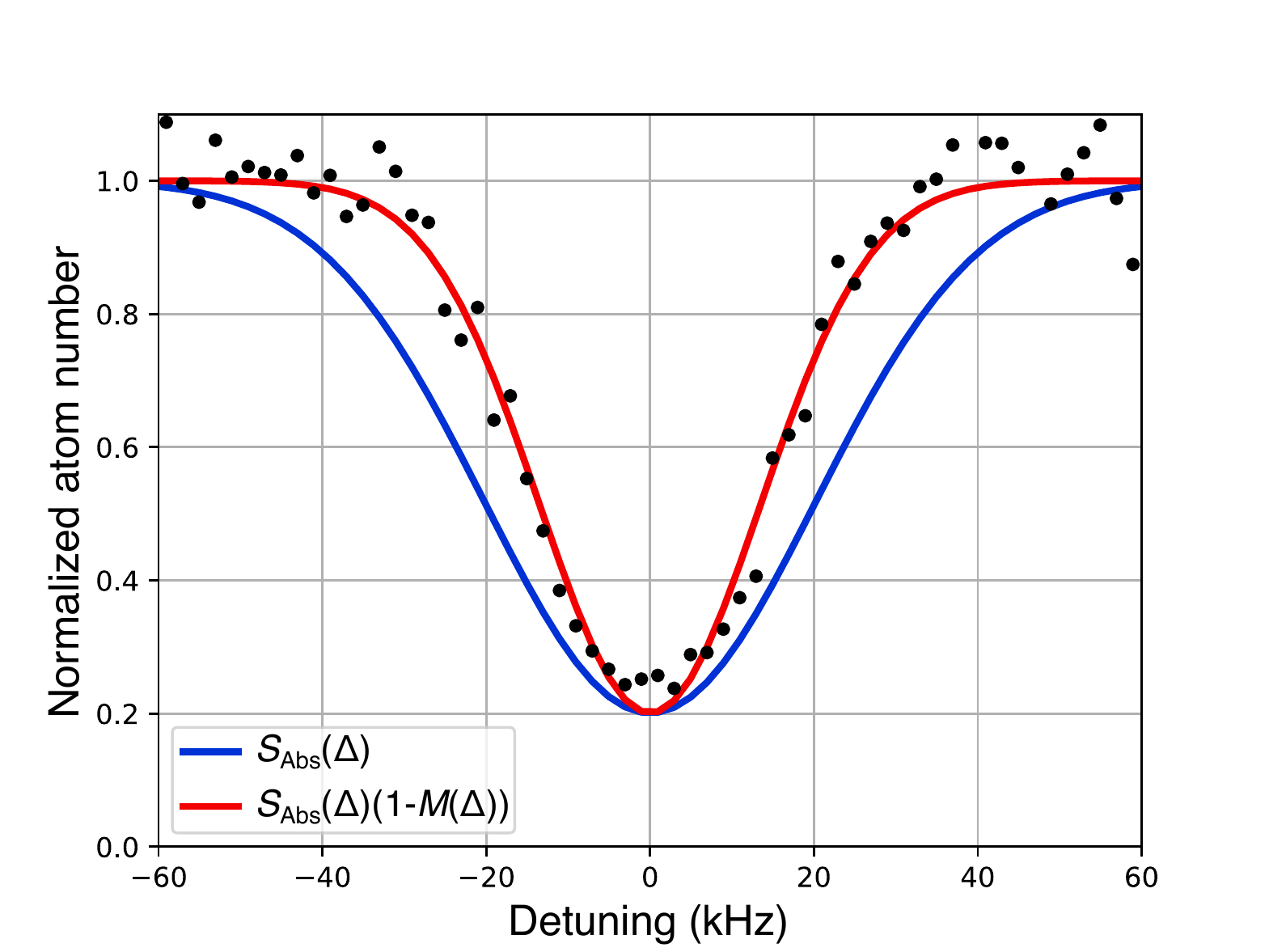}
\caption{Comparison of the observed spectrum (black dots) with the calculated Doppler broadened profile (solid blue line) and the narrowed profile based on Pauli-blocking when no depumper light is present (solid red line). 
}
\label{fig:spectrum}
\end{figure}

In Figure \ref{fig:spectrum}, our experimental observation of Pauli blocking on stimulated emission is shown for the $2^3\mathrm{S}_1\rightarrow 2^1\mathrm{S}_0$ transition at a temperature of $T/T_\mathrm{F}\simeq 0.55$. Compared to the calculated Doppler width (the blue line in Fig. \ref{fig:spectrum}), the transition is narrowed by a factor 0.75, clearly indicating the expected effect of Pauli blocking. Note that the observed linewidth is independent of the excitation time and amount of depletion, which was confirmed by simulations. This is related to  the low Rabi frequencies involved in the excitation and the absence of rethermalizing $s$-wave collisions in the ultracold Fermi gas.

From equation \eqref{eq:real_model}, for $\mathcal{M}(\omega)$ to modify the line profile significantly, the lifetime $\tau=1/\Gamma_0$ of the $\ket{e}$ internal state should be longer than $f_g/\Omega^2$. Therefore, the Pauli blockade effect can be suppressed by artificially decreasing the lifetime of state $\ket{e}$. This prevents excited atoms to be stimulated back to the $\ket{g}$ state, so that a Doppler broadened profile is retrieved. It effectively creates the same situation as previously demonstrated for helium in a \SI{1557}{nm} ODT \cite{Notermans2016} (blue-detuned dipole potential for the $2^1\mathrm{S}_0$ state), where the excited state $\ket{e}$ is expelled from the trap so that it cannot be stimulated back. Artificially decreasing the lifetime of the upper state thus provides an experimental way to verify the influence of Pauli blockade on stimulated emission.

We can effectively decrease the lifetime of the $2^1\mathrm{S}_0$ state by exciting it to the short-lived $4^1\mathrm{P}_1$ state with a \SI{397}{nm} "depumper" laser \cite{Martin1960, Guan2019}.
The relevant four-level system is depicted in Fig. \ref{fig:levels_flopping}(a). Any excitation to the $4^1\mathrm{P}_1$ state will result in a quick decay to the $1^1\mathrm{S}_0$ level since the Einstein A coefficient for this transition is 35 times higher than the one leading back to $2^1\mathrm{S}_0$ \cite{NIST2020}. Therefore, those atoms will be lost from the Fermi gas. By varying the intensity of the \SI{397}{nm} depumper, we are able to adjust the effective lifetime of the $2^1\mathrm{S}_0$ state ($1/\Gamma_0$ in  equation \eqref{eq:real_model}) and thus change the effect of the modification factor $\mathcal{M}(\omega)$ on the linewidth.

\begin{figure}
\center
\includegraphics[scale=1]{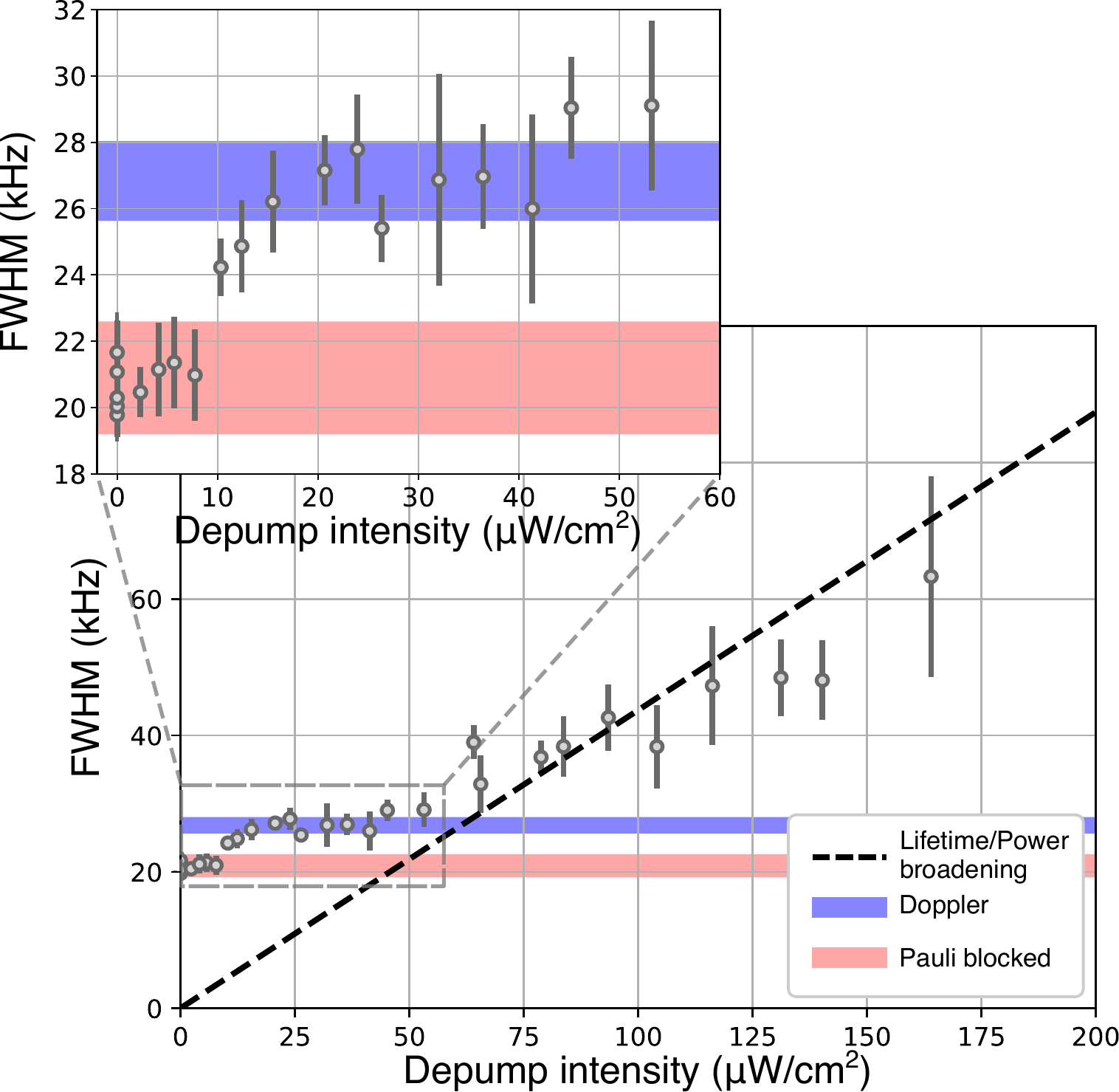}
\caption{Observed FWHM versus depumper intensity. The regimes where narrowing happens as a consequence of Pauli blocking and where the profile is fully Doppler-dominated are seen as the first and second plateau respectively. Their expected values from the parameters of the Fermi gases are displayed as red and blue areas respectively. The dashed black line shows the expected FWHM based on only lifetime and power broadening obtained from numerically solving the optical Bloch equations describing the evolution of a four level system for a single atom. The inset provides a zoom of the Pauli blocking and Doppler broadened regime. }
\label{fig:measurement396}
\end{figure}

Fig. \ref{fig:measurement396} shows the observed full width at half maximum (FWHM) of the spectroscopic lines as a function of depumper laser power. The two shaded regions indicate the expected linewidths, for the two mentioned regimes, calculated with expression \eqref{eq:general_S} based on averaged thermodynamical values of the Fermi gases (see the Methods section).
In order to isolate any trivial broadening due to the depumper (lifetime or power broadening), we numerically solve the optical Bloch equations averaged over the Fermi-Dirac distribution of the gas for the level scheme shown in Fig. \ref{fig:levels_flopping}(a) (see the Methods section for details). The predicted FWHM obtained using this procedure as a function of depumper intensity is shown as a dashed line. At intensities above approximately \SI{60}{\mu W/cm^2}, the width is dominated by single-atom lifetime broadening due to the depumper beam. Below that intensity, the quantum statistical effects can be distinguished in two clear regimes. The first one appears when no depumper light (or a small intensity) is applied. This case corresponds to the first plateau (FWHM$\ \sim\ $\SI{21}{kHz}) seen in Fig. \ref{fig:measurement396}, with a sub-Doppler linewidth due to Pauli blockade. 

As the intensity of the depumper beam is increased, the lifetime of the singlet state becomes shorter so that eventually stimulated emission is suppressed, and therefore the influence of Pauli blockade disappears.
This regime constitutes the second plateau (FWHM$\ \sim\ $\SI{27}{kHz}) seen in Figure \ref{fig:measurement396}. The expected values of the two plateaus are well predicted by expression \eqref{eq:general_S} considering an effective lifetime for the excited state. In order to also predict the threshold intensity where the transition happens between the two regimes, one would have to include the full effect of the excitation laser linewidth in the model. As mentioned before, this would further complicate the calculations significantly, and was therefore not pursued.

\begin{figure}
\center
\includegraphics[scale=1]{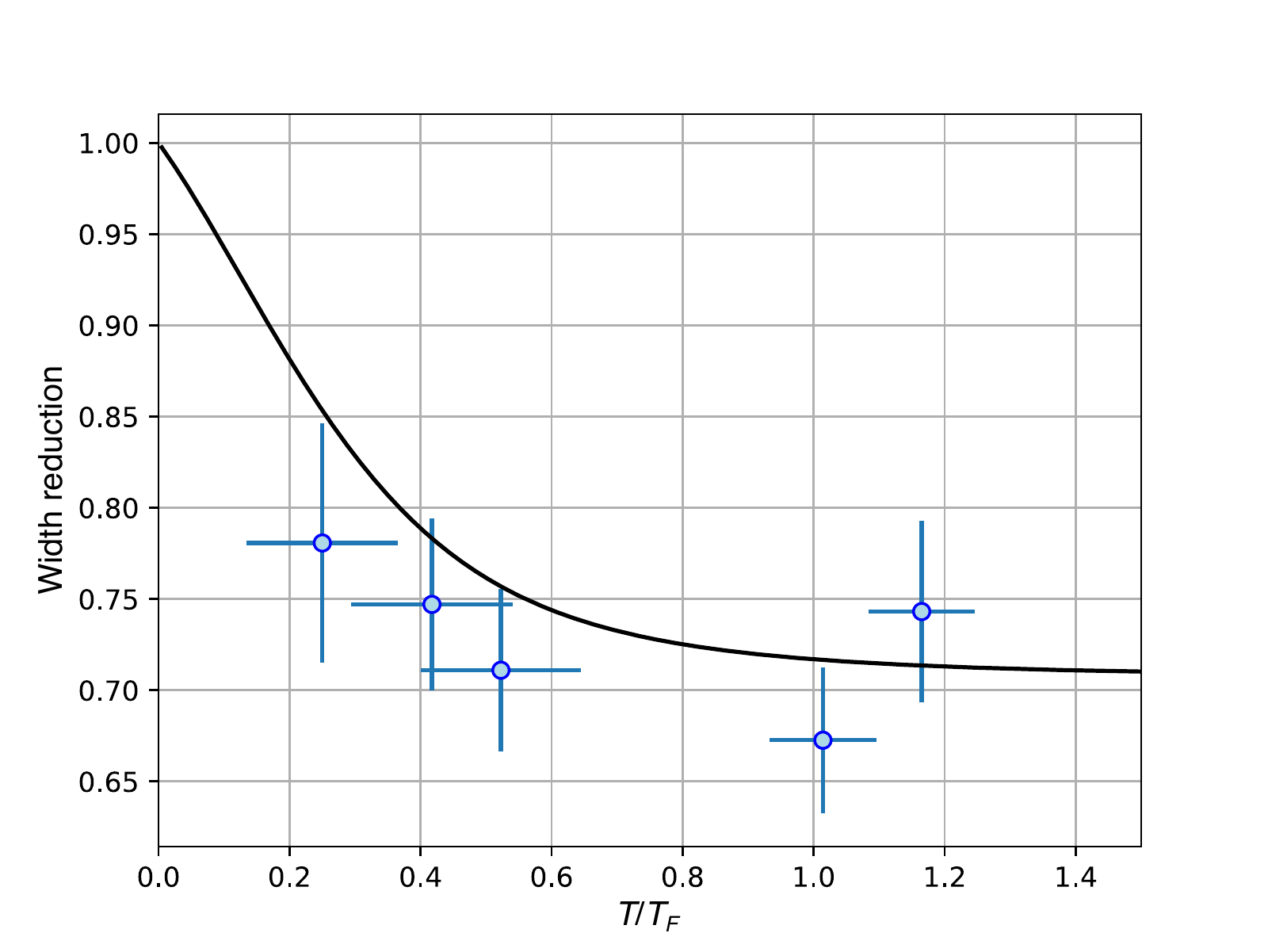}
\caption{Reduction of the FWHM due to the effect of Pauli blocking of stimulated emission as a function of the temperature $T/T_F$. The solid black curve corresponds to the ratio of the expected widths of $\mathcal{S}(\omega)$ calculated using expression \eqref{eq:general_S}, and $\mathcal{S}_\mathrm{Absorption}(\omega)$. Blue points represent the values of the ratio of the two plateaus averaged over intervals of $T/T_F$ values.}
\label{fig:ratio_T_TF}
\end{figure}

By varying our experimental cooling and trapping conditions, we can change the parameters of the produced Fermi gases over a range of temperatures and chemical potentials. It is then possible to construct a universal curve displaying the linewidth reduction factor as a function of temperature. This is shown in Figure \ref{fig:ratio_T_TF}.
In the limit of zero temperature, all states below the Fermi energy are occupied (except one hole from excitation), hence it is very improbable for excited atoms to be stimulated back into the lower electronic state, resulting in a Doppler excitation profile. As the temperature of the Fermi gas increases, the availability of holes around the Fermi energy increases too and stimulated emission becomes possible, leading to an increased narrowing of the line profile. When $T \gg T_F$, the quantum statistical nature of the atoms does not play a role anymore in the expression of the rates \eqref{eq:rate_ge} and \eqref{eq:rate_eg}, and the Doppler profile should be retrieved again (this is out of the validity range of the presented theoretical approach). As can be seen in Figure \ref{fig:ratio_T_TF}, the reduction of the actual FWHM compared to the Doppler linewidth becomes stronger as $T/T_F$ increases until it reaches a reduction by a factor $\sim 0.7$ when the temperature is comparable to the Fermi temperature. 
Although it was not possible to achieve temperatures lower than $T\simeq 0.25\  T_F$ to measure into the deeply degenerate regime where the availability of holes becomes negligible, the experimental values consistently show a significant narrowing compared to the Doppler width, with a good agreement to the theoretical curve over the full range of experimental parameters.

In contrast to what was observed by \cite{Deb2021, Sanner2021, Margalit2021}, our study shows that Pauli blockade can affect coherent processes in specific situations where quantum exchange symmetry plays a role. Interestingly, a broader linewidth of the excitation laser results in this case in a narrowing of the spectroscopic linewidth. Even though we are experimentally limited to temperatures down to $T/T_F\simeq 0.25$, our narrowed spectra mostly reflect the low momentum atoms, so we nevertheless achieve linewidths which are otherwise only reachable for highly degenerate gases.
In a broader context, our observation of Pauli blockade of coherent processes could be of interest in the context of cooling of fermionic samples.
Additionally, the mechanism shows similarities with conduction phenomena in semi-conductor materials and could be used to perform quantum simulations of such materials, or be of great interest for quantum logic and information processing with fermionic species.
   
In conclusion, we observed a signature of Pauli blockade in a coherently driven system, by the means of spectroscopy of a doubly-forbidden transition in an ultracold Fermi gas of metastable $^3\mathrm{He}$.
Due to a dependence of this Pauli blockade effect on the laser detuning, we observe reduced spectroscopic linewidths compared to the expected Doppler profiles arising from the finite momentum distribution of the gas. We show that, when actively reducing the lifetime of the excited state, we prevent all stimulated emission events and eliminate the Pauli blockade effect, by which the Doppler width is retrieved. The measured data are in excellent agreement over a range of parameters with the expectations from the model, which only requires the input of thermodynamic and laser parameters. In the context of precision spectroscopy, it proves to be a useful feature as it helps improving the accuracy of the determination of the transition frequency without inducing any additional shift. In a broader perspective, the narrowing effect showcases Pauli blockade of stimulated emission. This influence of quantum statistics on a coherently driven system shows great potential as a tool for applications with many-body physics and quantum information. 

\begin{methods}

\subsection{Experimental sequence.}
A beam of $^4\mathrm{He}$ and $^3\mathrm{He}$ in the metastable $2^3\mathrm{S}_1$ state is produced using a DC discharge. The beam of atoms is collimated before entering a \SI{2}{m} long Zeeman slower and subsequently captured by a 3D magneto-optical trap (MOT), producing samples at temperatures around \SI{600}{\mu K}. The MOT is loaded with a mixture of the two isotopes for \SI{5}{s}. The atomic cloud is then transferred to a cloverleaf magnetic trap where it is polarized to the $m_J=+1$ ($^4\mathrm{He}$) or the $m_F=+3/2$ ($^3\mathrm{He}$) spin states. The obtained spin-stretched cloud is then 1D-Doppler cooled for \SI{3}{s}, which lowers the temperature to about \SI{110}{\mu K}. Forced evaporative cooling is then performed on the mixture of the two isotopes by ramping down the frequency of an RF-knife from \SI{50}{MHz} to \SI{3}{MHz} in \SI{3}{s}. Collisions between the two isotopes allow rethermalization of the $^3\mathrm{He}$ component through sympathetic cooling. At the final RF frequency, depending on the value of the offset magnetic field produced by the trap coils, all $^4\mathrm{He}$ has evaporated and only a Fermi gas of $^3\mathrm{He}$ is left in the magnetic trap. The magic wavelength ODT beams, which are generated by doubling \SI{640}{nm} light produced by sum-frequency mixing of two beams at \SI{1557}{nm} and \SI{1086}{nm}, are then turned on before the confining magnetic field is turned off adiabatically in \SI{50}{ms}. Detection of the cloud is performed by turning off the ODT beams and free fall of the atoms onto an MCP located \SI{17.5}{cm} below the trapping region. A time-of-flight profile is fitted to the recorded trace from which the thermodynamical parameters ($N$, $\mu$ and $T$) are extracted.

\subsection{Data acquisition and analysis.}
Before and after each measurement of a spectroscopic profile, a set of reference Fermi gases is measured in order to extract thermodynamical parameters ($N$, $\mu$ and $T$). These two sets are averaged in order to account for any drift or noise during the realization of the spectroscopy scan. Each spectroscopic measurement is obtained by holding a DFG in the ODT for \SI{3}{s} while it is exposed to excitation light, after which the same sequence is performed again with the probe light turned off in order to account for atom number fluctuations. A depletion of up to \SI{80}{\%} of the initial trapped atoms is achieved at the resonance frequency. While the spectroscopy laser is on, a constant magnetic field of \SI{4}{G} is applied to minimize depolarization of the sample. 
A spectrum is then built up by determining the number of atoms left in the gas for each laser frequency, normalized to the reference measurements where no light was present. For typical conditions, the spectral line shape we obtain can be well approximated and fitted with a Gaussian profile, from which we determine the FWHM of the line. The degeneracy of the initial Fermi gas can be varied by a combination of changing the sweeping rate of the RF knife frequency and changing the power of the ODT beams. The data shown in Figure \ref{fig:measurement396} have been acquired using \SI{300}{mW} of light in each beam of the crossed ODT and a trapped Fermi gas composed of about $1.5\times 10^5$ atoms at a temperature of $\sim$\SI{100}{nK}. These data are combined in Figure \ref{fig:ratio_T_TF} with other sets of data, obtained with DFG composed of $10^5$ to $10^6$ atoms and temperatures ranging from \SI{100}{nK} to \SI{300}{nK}. 
Displayed as a function of depumper laser intensity, these data exhibit a similar behaviour showing two regimes as those in Figure \ref{fig:measurement396} (see Supplementary for details and figure). For each set, we determine the temperature $T/T_F$ through the relation:
\begin{equation}
\dfrac{T}{T_F} = \left( -6\ \mathrm{Li}_3(-\zeta) \right)^{-1/3},
\end{equation}
where $\mathrm{Li}_3(-\zeta)$ is the trilogarithm function and $\zeta = e^{\beta\mu}$ is the fugacity of the gas. The measurements are divided into two categories: below the depumping intensity threshold where the stimulated emission is relevant, and above this threshold where the stimulated emission is eliminated. We discard lifetime-broadened measurements from the analysis shown in Figure \ref{fig:ratio_T_TF}. The two sets of data obtained this way are then binned based on their values of $T/T_F$ and a value of the linewidth is obtained by averaging over the range of the binning interval. Finally, the ratio between the widths of the profiles, with and without the contribution of the Pauli blocked stimulated emission, is calculated and errors are propagated.

\subsection{Generation of the depumper light.}
The light resonant with the $2^1\mathrm{S}_0 \rightarrow 4^1\mathrm{P}_1$ transition at \SI{396.603}{nm} is generated by an external cavity laser diode locked to a saturated absorption signal from a discharge cell of $^3\mathrm{He}$. The frequency is also continuously monitored using a High Finesse WS-U 30 wavelength meter to prevent any drift during the rather long ($\sim$ 1 hour) acquisition of a spectroscopic scan. 

\subsection{Estimation of the lifetime broadening.}
Lifetime broadening due to the coupling of the $2^1\mathrm{S}_0$ and the $4^1\mathrm{P}_1$ states is estimated by solving a set of 16 optical Bloch equations (accounting for the 4 populations and 12 coherences of the 4 internal states) describing the dynamics of the $1^1\mathrm{S}_0$, $2^3\mathrm{S}_1$, $2^1\mathrm{S}_0$ and $4^1\mathrm{P}_1$ as shown on Figure \ref{fig:levels_flopping}(a) (see for example Ref. \cite{Rooijakkers1997}). All values of the Einstein coefficients are taken from Ref. \cite{NIST2020}. Such a set of equations can be written in the form:
\begin{equation}
\dot{\mathbf{\rho}}=\mathbf{M}\cdot \mathbf{\rho},
\end{equation}
where the density matrix $\mathbf{\rho}$ is flattened into a 16 elements column vector and $\mathbf{M}$ is a $16\times 16$ matrix containing the couplings. The solution is obtained by propagating the initial density matrix by numerically computing the expression:
\begin{equation}
\mathbf{\rho}(t) = e^{\mathbf{M}t}\cdot \mathbf{\rho}(0),
\label{eq:4_levels}
\end{equation}
where initially the only non-zero element is the one describing a population in the metastable triplet state (equal to 1). The $2^3\mathrm{S}_1$ population at time $t=\ $\SI{3}{s} is evaluated over a range of \SI{500}{kHz} around the resonance frequency to build up a spectrum, from which the FWHM is extracted. 
Since our modelling describes only single atom dynamics, the coupling between the metastable triplet and singlet states inserted in equation \eqref{eq:4_levels} is obtained by averaging the Rabi frequency over the Fermi-Dirac distribution following a similar approach as in Ref. \cite{Campbell2009} (see the Supplementary Material). 

\subsection{Data availability.}
The data that support the plots within this paper and other findings of this study are available from the corresponding author upon reasonable request.

\end{methods}

\bibliography{main}

\begin{addendum}
 \item We are greatly indebted to the late Wim Vassen for initiating and leading the metastable helium research at our university for many years. To honour his contributions we dedicate this article to him. We would also like to thank NWO (Netherlands Organisation for Scientific Research) for funding through the Program "The mysterious size of the proton" (16MYSTP) and Projectruimte grant 680-91-108. We thank Rob Kortekaas for technical support.
 \item [Author contributions] R.J., Y.v.d.W. and K.S. performed the measurements. R.J. and Y.v.d.W. analyzed the data. R.J. developed the theoretical model and performed the numerical simulations. H.L.B. and K.S.E.E. supervised the project. All authors contributed to the writing of the manuscript and the discussion of the results. 
 \item[Competing Interests] The authors declare no
competing interests.
 \item[Correspondence] Correspondence and requests for materials
should be addressed to R. Jannin (email: r.jannin@vu.nl) or K.S.E. Eikema (email: k.s.e.eikema@vu.nl) .
\end{addendum}

\newpage

\newpage

\newpage

\newpage

\newpage

\end{document}


\maketitle

\section{\textbf{Calculation of the transition rates and derivation of the line profile}}

As explained in the main text, we consider a Fermi gas in an internal state $\ket{g}$ confined in a harmonic potential. This trapping potential has an angular frequency $\omega_g$ along its axial direction. State $\ket{g}$ is coupled at time $t=0$ to the excited state $\ket{e}$ by resonant light with a wavenumber $\kappa$ and angular frequency $\omega$. The energy difference between these two levels is $\hbar\omega_0$. The excitation laser beam propagates along the axial direction of the trapping potential (which we denote $\sslash$ whereas the radial direction is denoted $\perp$). The atoms in state $\ket{e}$ feel a harmonic potential with an axial angular frequency $\omega_e$, which in the particular case of a magic wavelength trap is $\omega_e = \omega_g$. The confinement potential is assumed to be weak compared to the recoil energy $\hbar\omega_\mathrm{rec}$ of the transition, such that the system is out of the Lamb-Dicke regime ($\eta\gg 1$, with the Lamb-Dicke parameter defined as $\eta = \sqrt{\omega_\mathrm{rec}/\omega_g}$ and the recoil energy $\hbar\omega_\mathrm{rec} = \hbar^2\kappa^2/2m$). For the transition of metastable helium considered in this study, the Lamb-Dicke parameter is $\eta\simeq 30$.

Consider first that the excitation light has a narrower spectral linewidth than the splitting in energy of the motional states. It is assumed to be tuned to a central frequency $\omega = \omega_0 + \ell_0\omega_g$. The system can be described by the following Hamiltonian in the interaction picture:
\begin{equation}
H = \sum_n \dfrac{\hbar\Omega_{n,n+\ell_0}}{2}\left( \hat{e}_{n+\ell_0}^\dagger \hat{g}_{n} + \hat{e}_{n+\ell_0} \hat{g}_{n}^\dagger \right),
\end{equation}
where $\hat{i}_n^\dagger$ and $\hat{i}_n$ represent the fermionic creation and annihilation operators of an atom in internal state $\ket{i}$ and motional state $\ket{n}$, and the Rabi angular frequencies are given by \cite{Leibfried2003}:
\begin{equation}
\Omega_{n, n + \ell_0} = \Omega e^{-\eta^2/2}\eta^{\left|\ell_0\right|}\sqrt{\dfrac{n_< !}{n_> !}} L_{n_<}^{\left|\ell_0\right|}(\eta^2),
\label{eq:rabi_freq}
\end{equation}
where $n_<$ ($n_>$) is the lesser (greater) of $n$ and $n + \ell_0$. The evolution of the system is straightforward and leads to independent Rabi oscillations between pairs of $\ket{g, n}$ and $\ket{e, n+\ell_0}$ states, as depicted on Figure 1(a) from the main text. These sets of transitions constitute the carrier transitions of the system. Since all these transitions couple pairs of states independently, no collective many-body effect is expected and each fermion composing the Fermi sea can be seen as a single particle.


Now consider the case where the excitation laser is broader than the energy splitting due to the confining potential. The previous Hamiltonian is modified and reads:
\begin{equation}
H = \sum_n\sum_{\delta\ell=-\infty}^{+\infty} a_{\delta\ell}(\ell_0)\dfrac{\hbar\Omega_{n,n+\ell_0+\delta\ell}}{2}\left( \hat{e}_{n+\ell_0+\delta\ell}^\dagger \hat{g}_{n} + \hat{g}_{n}^\dagger\hat{e}_{n+\ell_0+\delta\ell}  \right),
\label{eq:H_broad}
\end{equation}
where $a_{\delta\ell}(\ell_0)$ accounts for the spectral density of the excitation laser at frequency $\omega_0 + (\ell_0+\delta\ell)\omega_g$. We will use in the following the more compact and convenient expression:
\begin{equation}
H = \sum_{nm} H_{g_n e_m} \left( \hat{e}_{m}^\dagger \hat{g}_{n} + \hat{g}_{n}^\dagger \hat{e}_{m} \right),
\end{equation}
with $H_{g_n e_m}/\hbar = a_{m-n}\Omega_{g_n,e_m}/2$.
From this expression, it is clear that sideband transitions, as shown in Figure 1(b) from the main text, will be driven in addition to carrier transitions. Hence, the Pauli exclusion principle has to be taken into account to model the proper time evolution of the system. 

In our experiment, the internal state $\ket{e}$ ($2^1\mathrm{S}_0$ state) has a lifetime $\tau = 1/\Gamma_0$ of 20 ms before it decays to the ground state ($1^1\mathrm{S}_0$ state) which is not trapped by the dipole potential. Hence, we are only interested in the first-order transition rates between the states $\ket{g, n}$ and $\ket{e, m}$. First, we examine the case of the $\ket{g,n} \rightarrow \ket{e,m}$ transition. In the occupation number representation, the initial state involved in the transition reads $\ket{i} = \ket{n_{g_n}, n_{e_m}}$ and the final state is necessarily:
\begin{equation}
\ket{f}=\ket{n_{g_n}-1, n_{e_m}+1} = \dfrac{\hat{e}_m^\dagger \hat{g}_n}{\sqrt{n_{g_n}\left( 1-n_{e_m} \right)}}\ket{i}.
\end{equation}
The coupling matrix element then reads:
\begin{equation}
\bra{f}H\ket{i} = H_{g_n e_m} \dfrac{\bra{i}\hat{g}_n^\dagger \hat{e}_m \hat{e}_m^\dagger \hat{g}_n\ket{i}}{\sqrt{n_{g_n}\left( 1-n_{e_m} \right)}} = H_{g_n e_m}\sqrt{n_{g_n}\left( 1-n_{e_m} \right)},
\end{equation}
where we have used the anti-commutation of fermionic operators and from which we get the transition rate using Fermi's golden rule:
\begin{equation}
\Gamma_{g_n\rightarrow e_m} = \dfrac{2\pi}{\hbar^2\omega_g} \left| H_{g_n e_m} \right|^2n_{g_n} \left( 1-n_{e_m} \right),
\label{eq:rate_g_e}
\end{equation}
since the density of states for a harmonic oscillator is $\rho(\epsilon) = 1/\hbar\omega_g$.
A very similar reasoning gives the rate from $\ket{e,m}$ to $\ket{g,n}$:
\begin{equation}
\Gamma_{e_m \rightarrow g_n} = \dfrac{2\pi}{\hbar^2\omega_e} \left| H_{g_n e_m} \right|^2 n_{e_m} \left( 1-n_{g_n} \right).
\label{eq:rate_e_g}
\end{equation}
These expressions clearly show that the two transitions are Pauli-blocked and that the transition probability will depend on the occupation number of the final state.

In order to model our system, we consider a simple rate equation for the $\ket{e,m}$ state:
\begin{equation}
\dot{n}_{e_m} = \sum_n \Gamma_{g_n\rightarrow e_m} - \sum_n \Gamma_{e_m \rightarrow g_n} - \Gamma_0 n_{e_m},
\end{equation}
from which we determine the condition on the occupation number for a steady state to be reached, $\dot{n}_{e_m}=0$. Using the expressions for the rates \eqref{eq:rate_g_e} and \eqref{eq:rate_e_g} and transforming the last equation, we obtain:
\begin{equation}
n_{e_m} = \dfrac{\omega_e}{\omega_g}\dfrac{\sum_j \left| H_{g_j e_m} \right|^2 n_{g_j}}{\sum_j \left| H_{g_j e_m} \right|^2\left[ 1-n_{g_j}\left( 1-\omega_e/\omega_g \right) \right] + \hbar^2\omega_e \Gamma_0/2\pi},
\label{eq:n_e_steady}
\end{equation}
which, in the limit where $\omega_e/\omega_g\simeq 1$ (close to the magic wavelength condition), simplifies to:
\begin{equation}
n_{e_m} = \dfrac{\sum_j \left| H_{g_j e_m} \right|^2 n_{g_j}}{\sum_j \left| H_{g_j e_m} \right|^2 + \hbar^2\omega_g \Gamma_0/2\pi}.
\label{eq:n_e_steady}
\end{equation}
The occupation number of the state $\ket{g,n}$ after a small excitation time $\Delta t$ (compared to the excitation rate) reads:
\begin{align}
n_{g_n}(\Delta t) \simeq & \ n_{g_n}(0) - \dfrac{2\pi}{\hbar^2}\dfrac{\Delta t}{\omega_g} \sum_m \left| H_{g_n e_m} \right|^2 n_{g_n}(0) \nonumber\\
& + \dfrac{2\pi}{\hbar^2}\dfrac{\Delta t}{\omega_g} \sum_m \left| H_{g_n e_m} \right|^2 n_{e_m}(\Delta t)\left( 1-n_{g_n}(0) \right),
\label{eq:model_deriv_interm}
\end{align}
in which we can replace $n_{e_m}(\Delta t)$ by the expression \eqref{eq:n_e_steady}, assuming that $\Delta t$ is long enough for a steady state to have been reached ($\Delta t \gg 1/\Gamma_0$). We obtain:
\begin{align}
n_{g_n}(\Delta t) \simeq &\  n_{g_n}(0) - \dfrac{2\pi}{\hbar^2}\dfrac{\Delta t}{\omega_g} \sum_m \left[ \left| H_{g_n e_m} \right|^2 n_{g_n}(0) \right.\nonumber\\
& - \left.\dfrac{\sum_{j} \left| H_{g_n e_m}\right|^2\left|H_{g_j e_m} \right|^2 n_{g_j}(0)\left( 1-n_{g_n}(0) \right)}{\sum_j \left| H_{g_j e_m} \right|^2 + \hbar^2\omega_g \Gamma_0/2\pi} \right],
\label{eq:model_derived}
\end{align}
from which we get the total population in the internal $\ket{g}$ state as:
\begin{equation}
n_g(\Delta t) = \sum_n n_{g_n}(\Delta t).
\end{equation}
The last term of expression \eqref{eq:model_derived} is similar to a second-order Fermi golden rule but takes its origin in the fermionic nature of the atomic species. It describes an exchange of motional state within the $\ket{g}$ internal state through a combination of the absorption of a photon followed by a stimulated emission process, which is Pauli-blocked by the presence of the Fermi sea. From this term, it is clear that if the excitation light only couples the sets of states $\ket{g,n}$ and $\ket{e, n+l_0}$, then this term vanishes, the system does not exhibit Pauli blocking anymore and the previous expression reduces to a common Fermi's golden rule. 

We then write the shape of the depletion signal as:
\begin{equation}
\mathcal{S}(\omega) \propto \sum_{nm} \left|a_{m-n}\right|^2\left| \Omega_{g_n e_m} \right|^2 \left(n_{g_n}(0) -\dfrac{\sum_j \left|a_{m-j}\right|^2\left| \tilde{\Omega}_{g_j e_m}\right|^2  n_{g_j}(0)\left( 1-n_{g_n}(0) \right)}{\sum_j \left|a_{m-j}\right|^2\left| \tilde{\Omega}_{g_j e_m}\right|^2 + f_g \Gamma_0/\Omega^2} \right), 
\label{eq:real_model}
\end{equation}
with $f_g = \omega_g/2\pi$ and $\tilde{\Omega}_{g_j e_m}=\Omega_{g_j e_m}/\Omega$.

Because the strength of the couplings is modulated by the power spectrum of the excitation light, we expect that the main contributions to the last term come from the cases where $|n-j|$ is small. We also note that in the limit where $\Gamma_0\gg\Omega^2/f_g$ (which can be obtained by artificially decreasing the lifetime of the $\ket{e}$ state as explained in the main text), the second term in the last expression vanishes, thus cancelling the effect of the Pauli blockade. In normal operation of our experiment and without pumping to the $4^1\mathrm{P}_1$ state, we have $\tau = 20\ \mathrm{ms}$, a free Rabi frequency of about \SI{40}{Hz} and typical trap frequencies of $10\mathrm{\ to\ }30\ \mathrm{Hz}$, corresponding to a parameter $f_g\Gamma_0/\Omega^2$ ranging from $2.4\times 10^{-2}$ to $8\times 10^{-3}$, which can be neglected as assumed in the previous derivation.

\section{\textbf{Calculation of the line profile in the Thomas-Fermi approach}}

Because of the complexity of the dynamics, we restrict ourselves to a local density approach. Semiclassically, we consider the one-body unperturbed Hamiltonians for both internal states, which are given by:
\begin{equation}
H_{g, e}(\mathbf{r}, \mathbf{p}) = \dfrac{\mathbf{p}^2}{2m} + \dfrac{m}{2}\omega_{g, e}^2\left( \alpha_x^2 x^2 + \alpha_y^2 y^2 + \alpha_z^2 z^2 \right),
\label{eq:Hamilton}
\end{equation}
where the coefficients $\alpha_x$, $\alpha_y$ and $\alpha_z$ account for the  anisotropy of the trapping potential.
The interaction with light is treated as a perturbation which transfers momentum to the atoms, allowing to make use of the Thomas-Fermi approximation. In such an approximation, the density of states of the DFG in state $\ket{g}$ is given by:
\begin{equation}
\rho_g(\mathbf{r}, \mathbf{k}) = \dfrac{1}{(2\pi)^3}\dfrac{1}{1+\exp\left( \beta H_g(\mathbf{r}, \hbar\mathbf{k})-\beta\mu \right)},
\label{eq:FD_dist}
\end{equation}
where $\beta = 1/(k_B T)$ is the reciprocal temperature, and $k_B$ Boltzmann's constant. 

The line profile is found by rewriting the expression \eqref{eq:real_model} within the Thomas-Fermi approximation in the same spirit as \cite{Juzeliunas2001}. We assume for simplicity that the function defined in expression \eqref{eq:n_e_steady} is peaked at $j=n$ when its denominator is not dominated by $f_g\Gamma_0/\Omega^2$ (thereby neglecting the small change in phase space induced by stimulated emission). We then simplify it as:
\begin{equation}
\mathcal{S}(\omega) \underset{\sim}{\propto} \sum_{nm} \left|a_{m-n}\right|^2\left| \Omega_{g_n e_m} \right|^2 \left[n_{g_n}(0) - n_{g_n}(0)\left( 1-n_{g_n}(0) \right) \right], 
\end{equation} 
such that we can rewrite it in the semiclassical approach as:
\begin{equation}
\mathcal{S}(\omega) \propto  \int \mathrm{d}^3 \mathbf{r} \int \mathrm{d}^3 \mathbf{k}\ \left[ \rho_\mathrm{g}(\mathbf{r}, \mathbf{k}) - \rho_\mathrm{g}(\mathbf{r}, \mathbf{k})\left( 1-\rho_\mathrm{g}(\mathbf{r}, \mathbf{k}) \right) \right] \delta(\omega - \omega_{\mathbf{r}, \mathbf{k}}),
\end{equation}
with the resonance condition
\begin{equation}
\omega_{\mathbf{r}, \mathbf{k}} = \omega_0 + \omega_\mathrm{rec} + \dfrac{2\hbar\kappa k_\sslash}{2m} + \dfrac{m}{2\hbar}\left(\bar{\omega}_e^2-\bar{\omega}_g^2\right) r^2 .
\label{eq:resonance}
\end{equation}


The lineshape thus reads:
\begin{equation}
\mathcal{S}(\omega) \propto \mathcal{S}_\mathrm{Absorption}(\omega)\left( 1 - \mathcal{M}(\omega) \right),
\end{equation}
with the absorption profile given by \cite{Juzeliunas2001}:
\begin{equation}
\mathcal{S}_\mathrm{Absorption}(\omega) = \int \mathrm{d}^3 \mathbf{r} \int \mathrm{d}^3 \mathbf{k}\ \rho_\mathrm{g}(\mathbf{r}, \mathbf{k}) \delta(\omega - \omega_{\mathbf{r}, \mathbf{k}}),
\end{equation}
and the modification factor expressed as:
\begin{equation}
\mathcal{M}(\omega) = \dfrac{\int \mathrm{d}^3 \mathbf{r} \int \mathrm{d}^3 \mathbf{k}\ \rho_\mathrm{g}(\mathbf{r}, \mathbf{k}) \left( 1 - \rho_\mathrm{g}(\mathbf{r}, \mathbf{k})\right) \delta(\omega - \omega_{\mathbf{r}, \mathbf{k}}) }{\int \mathrm{d}^3 \mathbf{r} \int \mathrm{d}^3 \mathbf{k}\ \rho_\mathrm{g}(\mathbf{r}, \mathbf{k}) \delta(\omega - \omega_{\mathbf{r}, \mathbf{k}})}.
\end{equation}
It can be rewritten as:
\begin{equation}
\mathcal{M}(\omega) = \left[1 - \dfrac{\int\mathrm{d}^3\mathbf{r}\int\mathrm{d}^3\mathbf{k}\ \rho_\mathrm{g}^2(\mathbf{r}, \mathbf{k})\delta(\omega - \omega_{\mathbf{r}, \mathbf{k}})}{\int\mathrm{d}^3\mathbf{r}\int\mathrm{d}^3\mathbf{k}\ \rho_\mathrm{g}(\mathbf{r}, \mathbf{k})\delta(\omega - \omega_{\mathbf{r}, \mathbf{k}})}\right],
\end{equation}
in which the second term represents the Pauli blocking induced inhibition factor on stimulated emission. 

We follow the derivation from \cite{Juzeliunas2001} to obtain an expression for $\mathcal{S}_\mathrm{Absorption}(\omega)$. Making use of the spherical symmetry over space and cylindrical one over momenta, we rewrite the integrals as:
\begin{equation}
\int \mathrm{d}^3\mathbf{r} \int \mathrm{d}^3\mathbf{k} =  8\pi^2 \int^\infty_0 r^2 \mathrm{d}r \int_0^\infty k_\perp \mathrm{d}k_\perp \int_{-\infty}^\infty \mathrm{d}k_\sslash, 
\end{equation}
which yields after integration over the axial momenta $k_\sslash$:
\begin{equation}
\mathcal{S}_\mathrm{Absorption}(\omega) = \dfrac{m}{\pi\hbar \kappa} \int^\infty_0 r^2 \mathrm{d}r \int_0^\infty k_\perp \mathrm{d}k_\perp \dfrac{1}{1+\exp \left[ \dfrac{\beta m}{2} \bar{\omega}_g^2 r^2 + \dfrac{\beta \hbar}{4\omega_\mathrm{rec}}\tilde{\omega}^2 - \beta\mu + \dfrac{\beta \hbar^2}{2m} k_\perp^2 \right]},
\end{equation}
with
\begin{equation}
\tilde{\omega} = \omega - \omega_0 - \omega_\mathrm{rec} - \dfrac{m}{2\hbar}\left( \bar{\omega}_e^2 - \bar{\omega}_g^2 \right)r^2.
\end{equation}
Performing the integration over the transverse momenta $k_\perp$, performing the change of variable $R=r\bar{\omega}_g\sqrt{\beta m/2}$ and denoting the detuning $\Delta = \omega - \omega_0 - \omega_\mathrm{rec}$, we can obtain an expression for $\mathcal{S}_\mathrm{Absorption}(\Delta)$:
\begin{equation}
\mathcal{S}_\mathrm{Absorption}(\Delta) \propto \int_0^\infty \mathrm{d}R\ R^2\ \mathrm{ln}\left[ 1 + e^{-F(R, \Delta)} \right],
\end{equation}
with $F(R,\Delta) = R^2+\dfrac{\hbar\beta}{4\omega_\mathrm{rec}}\left( \Delta + \dfrac{m_\mathrm{ex}}{\beta\hbar} R^2 \right)^2 - \beta\mu$, where $m_\mathrm{ex} = 1 - \omega_e^2/\omega_g^2$ characterizes the deviation from the magic wavelength case, and where $R$ is only an integration variable. 

We follow the exact same procedure for the calculation of $\int\mathrm{d}^3\mathbf{r}\int\mathrm{d}^3\mathbf{k}\ \rho_\mathrm{g}^2(\mathbf{r}, \mathbf{k})\delta(\omega - \omega_{\mathbf{r}, \mathbf{k}})$, considering the magic wavelength case where $\omega_g = \omega_e$ and $m_\mathrm{ex} = 0$, and we find :
\begin{equation}
\mathcal{M}(\Delta) = \dfrac{1}{\mathcal{S}_\mathrm{Absorption}(\Delta)} \int_0^\infty\mathrm{d}R\ \dfrac{R^2}{1+e^{F(R, \Delta)}},
\end{equation} 
with $F(R, \Delta)$ which now reads:
\begin{equation}
F(R, \Delta) = R^2 + \dfrac{\hbar\beta}{4\omega_\mathrm{rec}}\Delta^2 - \beta\mu.
\end{equation}
Integrating $\mathcal{S}_\mathrm{Absorption}(\Delta)$ and $\mathcal{M}(\Delta)$ then gives:
\begin{align}
\mathcal{S}_\mathrm{Absorption}(\Delta) & \propto -\dfrac{\sqrt{\pi}}{4} \mathrm{Li}_{5/2}\left[ -\zeta \exp\left(-\dfrac{\hbar\beta}{4\omega_\mathrm{rec}}\Delta^2\right) \right],\\
\mathcal{M}(\Delta) & = \dfrac{ \mathrm{Li}_{3/2}\left[ -\zeta \exp\left(-\dfrac{\hbar\beta}{4\omega_\mathrm{rec}}\Delta^2\right) \right]}{\mathrm{Li}_{5/2}\left[ -\zeta \exp\left(-\dfrac{\hbar\beta}{4\omega_\mathrm{rec}}\Delta^2\right) \right]},
\end{align}
where the fugacity is $\zeta = e^{\beta\mu}$ and $\mathrm{Li}_n(z)$ denotes the $n$-th order polylogarithm function of $z$.

Hence, $\mathcal{S}(\Delta)$ becomes:
\begin{equation}
\mathcal{S}(\Delta) \propto \mathrm{Li}_{5/2}\left[-\zeta\, \exp\left(-\dfrac{\hbar \beta}{4 \omega_{rec}}\Delta^2\right)\right] - \mathrm{Li}_{3/2}\left[-\zeta\, \exp\left(-\dfrac{\hbar \beta}{4 \omega_{rec}}\Delta^2\right)\right].
\label{eq:two_polylogs}
\end{equation}
In case the lifetime of the $\ket{e}$ internal state $\tau$ becomes small compared to $f_g/\Omega^2$ (by applying sufficient intensity in the depump laser from the $2^1\mathrm{S}_0$ to the $4^1\mathrm{P}_1$ states as presented in the main text), the second term related to $\mathcal{M}(\omega)$, vanishes and we retrieve the non Pauli-blocked profile as derived in \cite{Juzeliunas2001} and experimentally verified in \cite{Notermans2016} (in that case, the condition $\tau \rightarrow 0$ was realized by the fact that the $\ket{e}$ state was expelled from the dipole trap by a blue-detuned harmonic potential).

\section{\textbf{Few-body numerical analysis}}

In order to verify our modelling, we numerically solved the master equation describing both the coherent excitation and the decoherence due to the decay to the $1^1\mathrm{S}_0$ state. The equation reads:
\begin{equation}
\dot{\rho}(t) = -\dfrac{i}{\hbar}\left[ \hat{H}, \rho(t) \right] + \sum_n \dfrac{1}{2}\left( 2\hat{C}_n\rho(t)\hat{C}_n^\dagger - \rho(t)\hat{C}_n^\dagger \hat{C}_n - \hat{C}_n^\dagger \hat{C}_n\rho(t) \right),
\label{eq:master_equation}
\end{equation}
where $\rho(t)$ represents the density matrix of the system at time $t$, and with the Hamiltonian given in expression \eqref{eq:H_broad} and the collapse operators defined as:
\begin{equation}
\hat{C}^\dagger_n = \sqrt{\Gamma_0}\ \hat{0}^\dagger_n \hat{e}_n\mathrm{\ and\ }
\hat{C}_n = \sqrt{\Gamma_0}\ \hat{e}^\dagger_n \hat{0}_n,
\end{equation}
the internal state $\ket{0}$ standing for the $1^1\mathrm{S}_0$ state and $\Gamma_0 = 1/\tau$. Since we are not interested in collective effects due to the decay processes, we modelled them as conserving the motional state while flipping the internal state from $\ket{e}$ to $\ket{0}$.
The solution $\rho(t)$ is numerically calculated using either an exact solving method \cite{Machnes2014, AmShallem2015} or a quantum Monte-Carlo one \cite{Molmer1993}. From the density matrix, the population of each manifold, $\ket{g}$, $\ket{e}$, or $\ket{0}$, is extracted according to:
\begin{equation}
\langle n_i(t)\rangle = \sum_n \mathrm{Tr}\left[ \hat{i}_n^\dagger \hat{i}_n \rho(t) \right]/N,
\end{equation}
$i$ standing for one of these three states and $N$ being the total number of particles. Due to the computational time increasing very fast, we limited our analysis to few-atoms systems not exceeding $N=3$.

\begin{figure}[ht]
\center
\includegraphics[scale=.5]{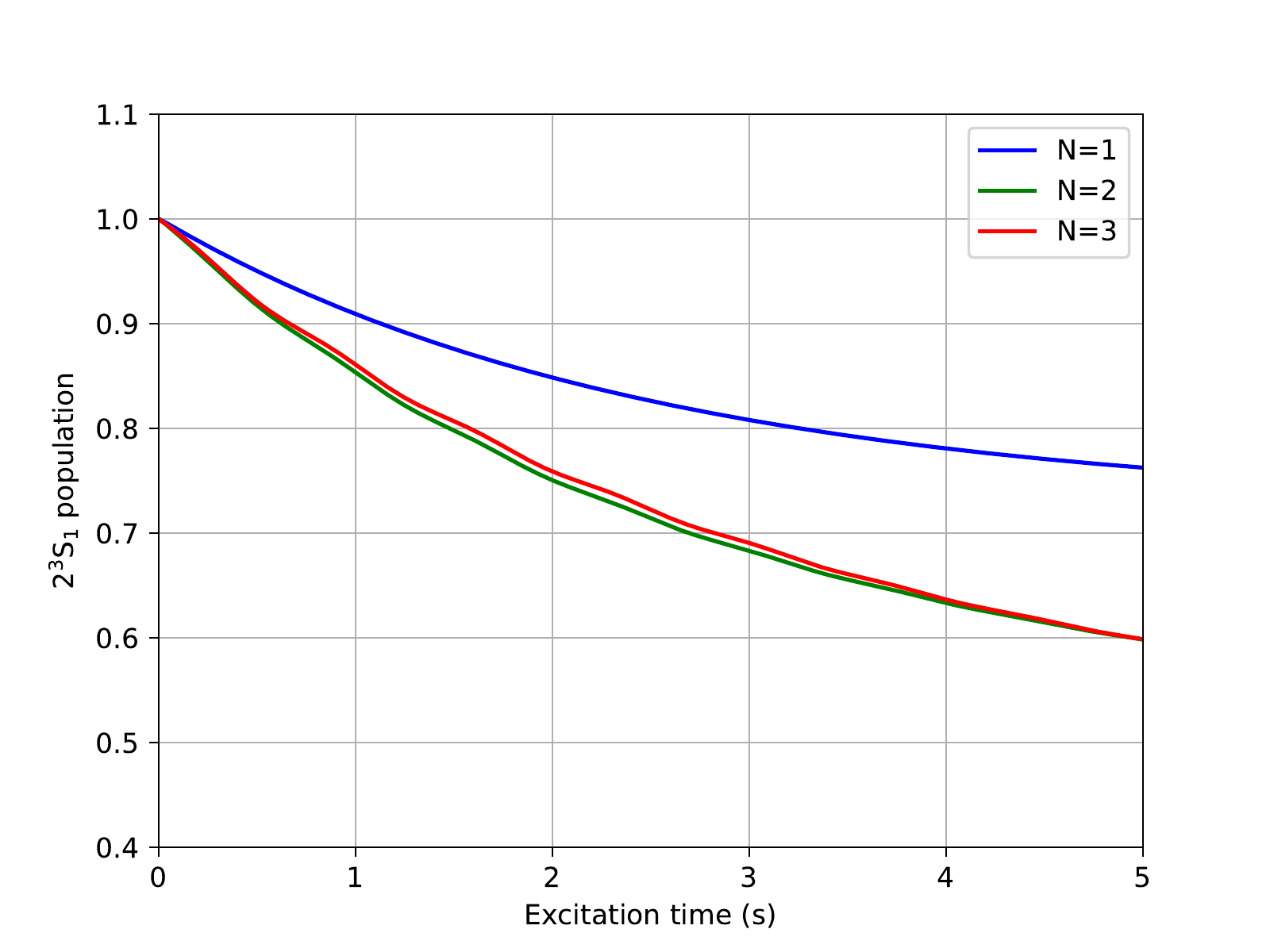}
\caption{Evolution of the population in the $2^3\mathrm{S}_1$ state upon excitation to the $2^1\mathrm{S}_0$ state for a system composed of 1 to 3 fermions obtained by solving the Master equation. In this example, the initial motional states of the atoms are centered around $n = 100$, in the energy range where the Pauli blockade changes the dynamics of the system.}
\label{fig:dynamics_g_state}
\end{figure}

A first thing that we confirmed is that no many-body effects can be observed if only carrier transitions are allowed and the Fermi gas can be treated as an ensemble of independent particles. In contrast, the dynamics is modified by the presence of the sideband couplings enabling cross-talk between the particles. The coherent dynamics exhibits Pauli blocking and can not be obtained by considering independent single-particles. Figure \ref{fig:dynamics_g_state} shows an example of such behaviour for different numbers of fermions in the system. Having more than a single atom results in an enhanced lower state depletion due to Pauli blockade.

\begin{figure}[ht]
\centerline{\includegraphics[scale=.5]{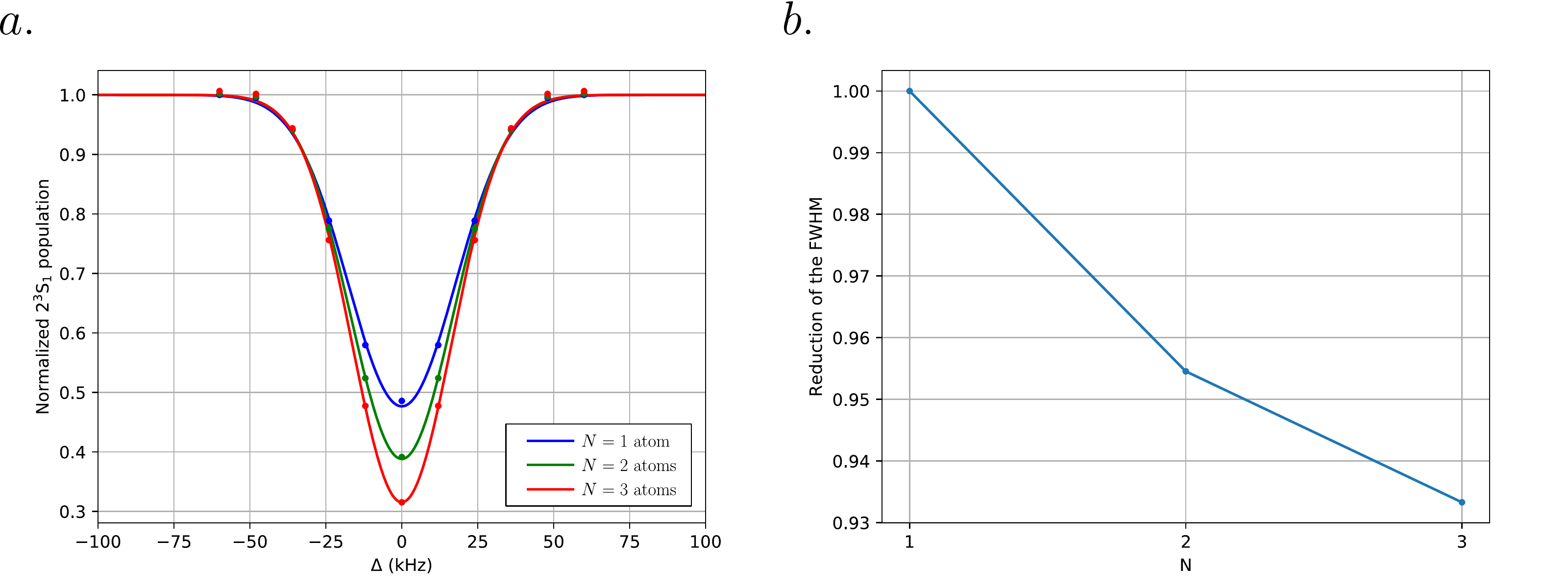}}
\caption{a. Simulation of line profiles with sets of nearest energy neighbours ranging from 1 to 3 atoms. b. Extracted FWHM as a function of the number of atoms. Increasing the number of fermions results in a decrease of the FWHM.}
\label{fig:few_body_lines}
\end{figure}

We then simulated line profiles by modelling the Fermi gas as an ensemble of sets of $N$ particles each of which is weighted by the Fermi-Dirac distribution. The remaining $\ket{g}$ population of each of those is calculated and added to obtain the total remaining population. The procedure is repeated for different detunings to simulate a line profile. Figure \ref{fig:few_body_lines}.a. shows the results of these simulations with increasing $N$ from 1 to 3. In our experiment, the Fermi gases are composed of $10^5-10^6$ atoms, hence we are only interested in a trend for the linewidth as it is not possible to simulate the full system. The excitation strength increases with $N$ and the profile narrows as shown in figure \ref{fig:few_body_lines}.b., displaying a similar behaviour as what we experimentally observe.

\section{\textbf{Averaging of the Rabi frequency}}
In order to estimate the average Rabi frequency that we use for solving the four level dynamics explained in the main text, we follow a similar approach as in Ref.\cite{Campbell2009}. We label the states with two indices, $n_\sslash$ and $n_\perp$, which represent the axial and radial directions respectively. With this notation, excitation light of angular frequency $\omega = \omega_0 + \omega_\mathrm{rec} + \ell\omega_\sslash$ couples only the sets of internal (electronic) and motional states $\ket{2^3\mathrm{S}_1}\ket{n_\sslash}$ and $\ket{2^1\mathrm{S}_0}\ket{n_\sslash+\ell}$ with a Rabi frequency $\Omega_{n_\sslash, n_\sslash + \ell}$ given by the equation \eqref{eq:rabi_freq}.
The averaged Rabi frequency is thus given by:
\begin{equation}
\langle \Omega \rangle = \dfrac{1}{N}\sum_{n_\sslash, n_\perp} f_\mathrm{FD}(n_\sslash, n_\perp)\ \Omega_{n_\sslash, n_\sslash + \ell},
\end{equation}
with the normalization $N = \sum_{n_\sslash, n_\perp}f_\mathrm{FD}(n_\sslash, n_\perp)$, and the Fermi-Dirac distribution given by:
\begin{equation}
f_\mathrm{FD}(n_\sslash, n_\perp)=\dfrac{1}{1+e^{-\beta\mu+\beta\hbar n_\sslash\omega_\sslash + \beta\hbar n_\perp\omega_\perp}}.
\end{equation}

\section{\textbf{Comparison of the behaviour of different datasets}}

\begin{figure}[ht]
\center
\includegraphics[scale=.75]{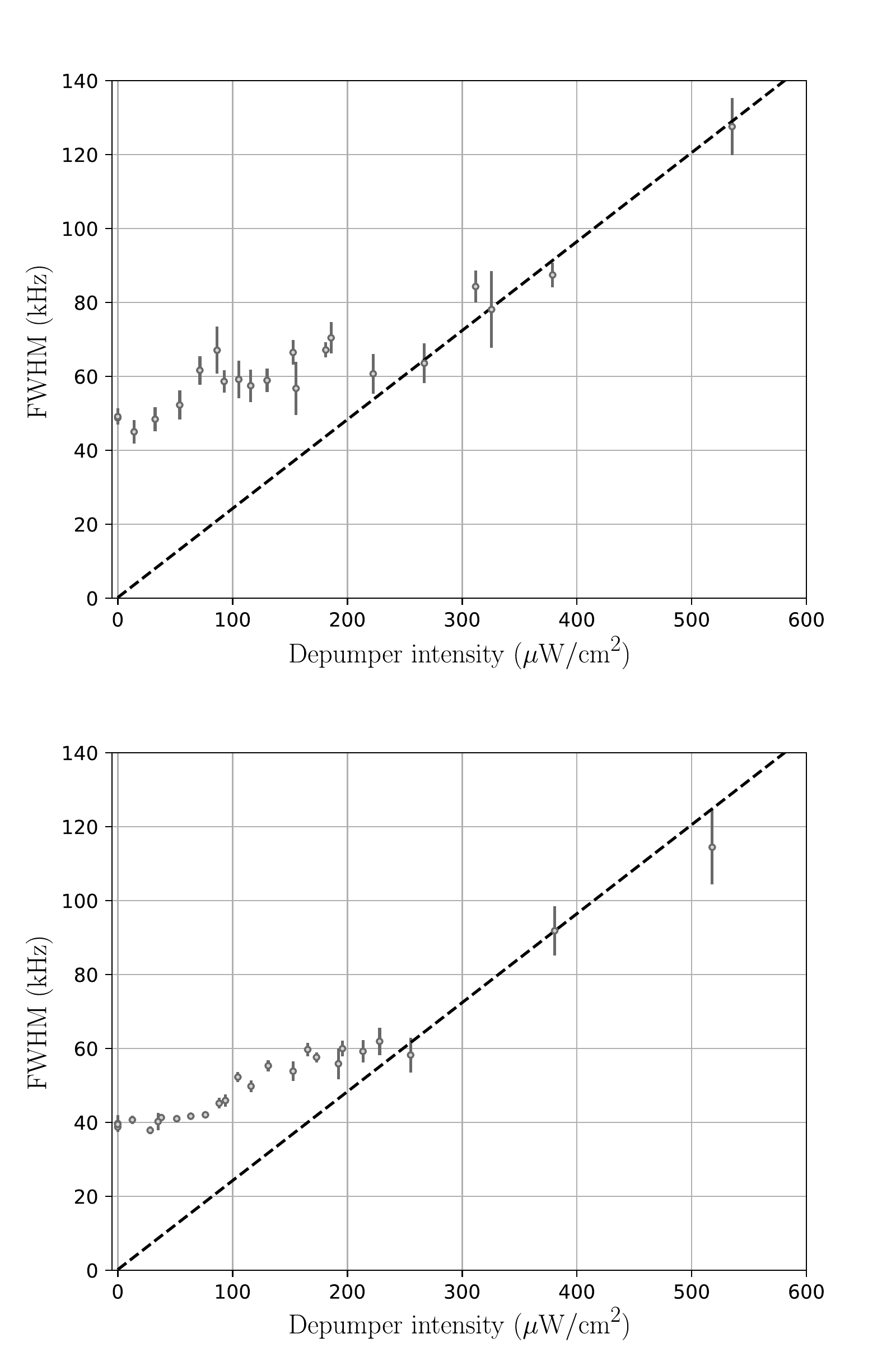}
\caption{Variation of the FWHM as a function of depumper laser intensity for deeper dipole traps than the data shown in the main text. Each of the two datasets were acquired with similar DFGs composed of $10^6$ atoms at a temperature of \SI{300}{nK} and exhibit the same behaviour as the dataset shown in the main text.}
\label{fig:750_data}

\end{figure}

As explained in the main text, we varied the intensity of the depumper beam for Fermi gases with different thermodynamical parameters. To achieve totally different $T/T_F$ values, we changed the intensity of the ODT beams. All the datasets we gathered this way exhibit the same behaviour as the one shown in the main text. Figure \ref{fig:750_data} shows a summary of two additional datasets obtained with different trap depth than Figure 3 of the main text.
The intensity value for which the transition between the two regimes happens is changed (also compared to the dataset shown in the main text) because of the interplay between the laser linewidth, the effective lifetime due to the pump to the $4^1\mathrm{P}_1$ state and the trap frequency. We did not take these data into account in the main text because it corresponds to an early implementation of the experimental procedure and the parameters were not well-controlled.

